\documentclass[runningheads]{llncs}

\usepackage{amsmath,amssymb,amsfonts}
\usepackage{graphicx}
\usepackage{booktabs}
\usepackage{multirow}
\usepackage{bm}
\usepackage{float}
\usepackage{placeins}
\usepackage[pagebackref,breaklinks,colorlinks,citecolor=blue]{hyperref}
\usepackage{cleveref}

\begin{document}
\emergencystretch=1.5em

\title{SAS-Net: Cross-Domain Image Registration as Inverse Rendering via Structure-Appearance Factorization}
\titlerunning{SAS-Net: Cross-Domain Registration as Inverse Rendering}

\author{Jiahao Qin\textsuperscript{*}}
\authorrunning{J. Qin}
\institute{Email: jiahao.qin19@gmail.com}

\maketitle
\let\thefootnote\relax\footnotetext{* Corresponding author.}

%==============================================================================
\begin{abstract}
\emergencystretch=1em
Cross-domain image registration requires aligning images acquired under heterogeneous imaging physics, where the classical brightness constancy assumption $I_m(\bm{x}) \approx I_f(T(\bm{x}))$ is fundamentally violated. We formulate this problem through an image formation model $I = \mathcal{R}(\bm{s}, \bm{a}) + \epsilon$, where each observation is generated by a \emph{rendering function} $\mathcal{R}$ acting on domain-invariant scene structure $\bm{s} \in \mathbb{R}^{C \times H \times W}$ and domain-specific appearance statistics $\bm{a} \in \mathbb{R}^{d}$. Registration then reduces to an \emph{inverse rendering} problem: given observations from two domains $\mathcal{D}_m$ and $\mathcal{D}_f$, recover the shared structure $\bm{s}_m$ and re-render it under the target appearance $\bm{a}_f$ to obtain the registered output $\hat{I}_{m \to f} = \mathcal{R}(\bm{s}_m, \bm{a}_f)$. We instantiate this framework as SAS-Net (Scene-Appearance Separation Network), where instance normalization implements the structure-appearance decomposition and Adaptive Instance Normalization (AdaIN) realizes the differentiable forward renderer. A scene consistency loss $\mathcal{L}_{\text{scene}}$ enforces geometric correspondence in the factorized latent space. Experiments on EuroSAT-Reg-256 (satellite remote sensing) and FIRE-Reg-256 (retinal fundus) demonstrate state-of-the-art performance across heterogeneous imaging domains. SAS-Net (3.35M parameters) achieves 89\,FPS on an RTX 5090 GPU.
Code: \url{https://github.com/D-ST-Sword/SAS-Net}.

\keywords{Inverse rendering \and Structure-appearance factorization \and Cross-domain registration \and Instance normalization \and AdaIN}
\end{abstract}

%==============================================================================
\section{Introduction}
\label{sec:introduction}
%==============================================================================

Let $I_m, I_f \in \mathbb{R}^{H \times W}$ denote a moving image and a fixed reference image acquired from two distinct imaging domains $\mathcal{D}_m$ and $\mathcal{D}_f$. Image registration seeks a spatial transformation $T: \mathbb{R}^2 \to \mathbb{R}^2$ such that $I_m \circ T \approx I_f$. Classical methods---SIFT~\cite{lowe2004distinctive}, Demons~\cite{vercauteren2009diffeomorphic}, optical flow~\cite{horn1981determining}, and SyN~\cite{avants2008symmetric}---assume brightness constancy $I_m(\bm{x}) \approx I_f(T(\bm{x}))$, which is violated when domain-specific imaging physics introduce systematic appearance differences between $\mathcal{D}_m$ and $\mathcal{D}_f$~\cite{chen2024survey,sotiras2013deformable}. Deep learning methods such as VoxelMorph~\cite{balakrishnan2019voxelmorph} and TransMorph~\cite{chen2021transmorph} learn deformation fields but similarly assume comparable intensity distributions across domains. Feature-level alignment strategies~\cite{qin2023atd,qin2024zoom} can mitigate modality gaps but do not explicitly model the physical image formation process.

\begin{figure}[!htb]
\centering
\includegraphics[width=\linewidth]{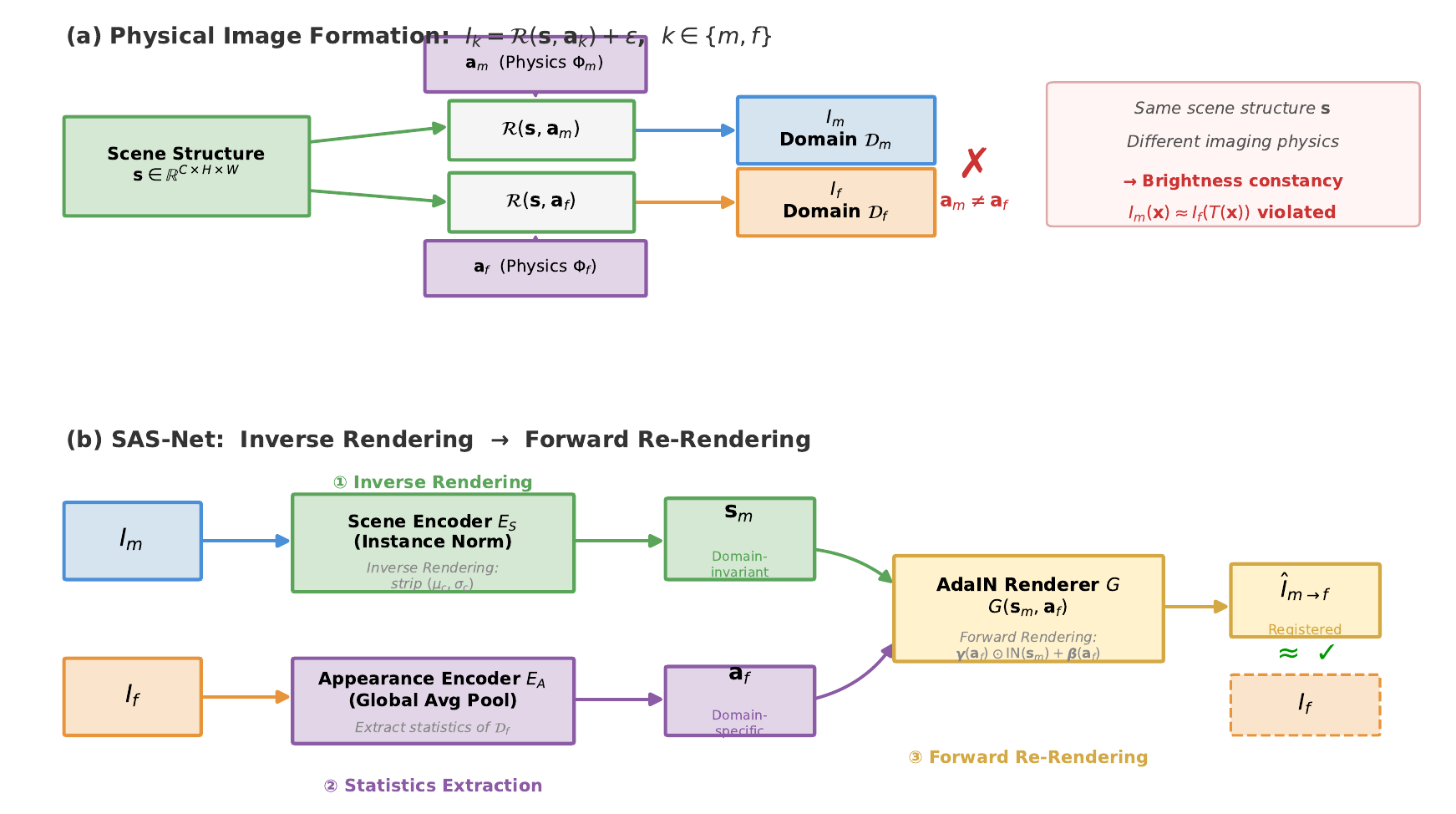}
\caption{Conceptual overview. (a)~\textbf{Physical image formation model}: The same scene structure $\bm{s}$ is rendered under two different imaging physics $(\bm{a}_m, \bm{a}_f)$, producing observations $I_m$ and $I_f$ in distinct domains $\mathcal{D}_m$ and $\mathcal{D}_f$. Direct pixel comparison fails because $\bm{a}_m \neq \bm{a}_f$, violating brightness constancy. (b)~\textbf{SAS-Net pipeline}: \textcircled{\small 1}~The scene encoder $E_S$ with instance normalization implements \emph{inverse rendering}---stripping domain-specific statistics $(\mu_c, \sigma_c)$ to recover domain-invariant structure $\bm{s}_m$. \textcircled{\small 2}~The appearance encoder $E_A$ extracts sufficient statistics $\bm{a}_f$ of the target domain via global average pooling. \textcircled{\small 3}~The AdaIN renderer implements \emph{forward re-rendering}---applying $\bm{a}_f$ to $\bm{s}_m$ via learned affine modulation $\bm{\gamma}(\bm{a}_f) \odot \mathrm{IN}(\bm{s}_m) + \bm{\beta}(\bm{a}_f)$---producing the registered output $\hat{I}_{m \to f} \approx I_f$.}
\label{fig:crossdomain_concept}
\end{figure}

\textbf{The cross-domain challenge.} As illustrated in \cref{fig:crossdomain_concept}, when the two images originate from different domains---e.g., satellite multispectral bands with distinct spectral response functions~\cite{helber2019eurosat}, or retinal fundus images acquired at different visits with varying illumination~\cite{hernandez2017fire}---the observed intensities are governed by domain-specific \emph{imaging transfer functions} $\Phi_k$. Each domain applies its own physics to the shared scene structure $\bm{s}$, producing observations $I_k = \mathcal{R}(\bm{s}, \bm{a}_k) + \epsilon_k$ where $\bm{a}_m \neq \bm{a}_f$. This coupling between appearance variation and geometric misalignment renders conventional registration ill-posed~\cite{chen2024survey}. Domain-invariant registration via disentangled representations~\cite{qin2026domainreg} and position-encoded temporal attention~\cite{wang2026gpereg} have shown promise, while progressive refinement strategies~\cite{qin2026pcreg} achieve coarse-to-fine alignment, and uncertainty-aware learning~\cite{qin2025dual} improves robustness under distributional shift. However, jointly correcting domain shift and geometric misalignment within a principled physical framework remains an open challenge.

\textbf{Key insight: registration as inverse rendering.} We observe that the cross-domain registration problem admits a natural \emph{inverse rendering} interpretation (\cref{fig:crossdomain_concept}b). If each image is generated by a forward rendering function $\mathcal{R}$ acting on two independent factors---domain-invariant scene structure and domain-specific appearance---then registration reduces to: (1) \emph{inverting} the rendering to recover the shared structure, and (2) \emph{re-rendering} the moving scene under the target domain's appearance. This factorization eliminates the need for explicit deformation field estimation, and the re-rendered output can further serve as input to energy-based reconstruction frameworks~\cite{wang2026dcer} for downstream compression tasks.

We instantiate this framework as SAS-Net and make the following contributions:
\begin{enumerate}
    \item We formalize cross-domain registration as an inverse rendering problem under the image formation model $I = \mathcal{R}(\bm{s}, \bm{a}) + \epsilon$, providing a principled physical foundation for structure-appearance factorization.
    \item We show that instance normalization and AdaIN jointly implement the inverse-forward rendering pipeline: IN extracts domain-invariant structure by removing appearance statistics, while AdaIN re-renders structure under target appearance via learned affine modulation.
    \item We demonstrate cross-domain generalization on EuroSAT-Reg-256 (satellite remote sensing) and FIRE-Reg-256 (retinal fundus), with 89\,FPS real-time capability, surpassing deformation-based methods under domain shift~\cite{qin2025bcpmjrs}.
\end{enumerate}

%==============================================================================
\section{Related Work}
\label{sec:related}
%==============================================================================

\paragraph{Deformable image registration.}
Classical approaches include diffeomorphic algorithms like Demons~\cite{thirion1998image} and SyN~\cite{avants2008symmetric}. Deep learning methods such as VoxelMorph~\cite{balakrishnan2019voxelmorph} and TransMorph~\cite{chen2021transmorph} predict deformation fields, while SynthMorph~\cite{hoffmann2021synthmorph} improves contrast robustness via synthetic training. Despite advances, existing methods assume globally comparable intensity distributions---an assumption violated under domain shift.

\paragraph{Image-to-image translation and disentangled registration.}
CycleGAN~\cite{zhu2017unpaired} enables unpaired translation through cycle consistency, while MUNIT~\cite{huang2018multimodal} and DRIT~\cite{lee2018diverse} disentangle content from appearance but do not enforce spatial alignment. Qin and Wang~\cite{qin2026domainreg} propose scene-appearance disentanglement for cross-domain registration, and Wang and Qin~\cite{wang2026gpereg} extend this with position-encoded temporal attention for sequential acquisitions. Progressive refinement approaches~\cite{qin2026pcreg} achieve high-fidelity registration through coarse-to-fine contrast guidance. The proposed SAS-Net differs by formulating registration as inverse rendering under an explicit physical image formation model, providing a principled separation between structure extraction and appearance transfer.

%==============================================================================
\section{Proposed Method: SAS-Net}
\label{sec:method}
%==============================================================================

\subsection{Problem Formulation: Image Formation under Heterogeneous Domains}
\label{sec:formulation}

\paragraph{Physical image formation model.}
Consider two imaging domains $\mathcal{D}_m$ and $\mathcal{D}_f$ characterized by distinct imaging transfer functions. Each domain captures the same underlying scene but under different physical conditions (e.g., spectral response, illumination, sensor noise). We model the image formation process as:
\begin{equation}
I_k(\bm{x}) = \mathcal{R}\bigl(\bm{s}(\bm{x}),\; \bm{a}_k\bigr) + \epsilon_k, \quad k \in \{m, f\},
\label{eq:forward_model}
\end{equation}
where $\bm{s}(\bm{x}) \in \mathbb{R}^{C}$ is the domain-invariant scene structure at spatial location $\bm{x} \in \Omega \subset \mathbb{R}^2$, $\bm{a}_k \in \mathbb{R}^{d}$ is a compact representation of domain-specific appearance (capturing global intensity statistics such as mean, contrast, and spectral response), $\mathcal{R}: \mathbb{R}^C \times \mathbb{R}^d \to \mathbb{R}$ is a differentiable rendering function, and $\epsilon_k$ is domain-specific noise.

\paragraph{Why brightness constancy fails.}
Classical registration assumes $I_m(\bm{x}) \approx I_f(T(\bm{x}))$, which implicitly requires $\bm{a}_m = \bm{a}_f$. When this condition is violated ($\bm{a}_m \neq \bm{a}_f$), the residual $\|I_m - I_f \circ T^{-1}\|$ contains both geometric misalignment \emph{and} appearance discrepancy, making it an unreliable registration objective.

\paragraph{Registration as inverse rendering.}
Under \cref{eq:forward_model}, cross-domain registration decomposes into two sub-problems:
\begin{enumerate}
    \item \textbf{Inverse rendering} (analysis): recover $\bm{s}_m, \bm{s}_f$ and $\bm{a}_m, \bm{a}_f$ from the observed images $I_m, I_f$;
    \item \textbf{Forward re-rendering} (synthesis): generate the registered output by rendering the moving scene structure under the fixed domain's appearance:
\end{enumerate}
\begin{equation}
\hat{I}_{m \to f} = \mathcal{R}(\bm{s}_m, \bm{a}_f).
\label{eq:cross_recon}
\end{equation}
If the factorization is \emph{identifiable}---i.e., $\bm{s}$ captures only geometry and $\bm{a}$ captures only appearance---then $\hat{I}_{m \to f} \approx I_f$ whenever $\bm{s}_m \approx \bm{s}_f$, achieving registration without deformation field estimation.

\begin{figure}[!htb]
\centering
\includegraphics[width=\linewidth]{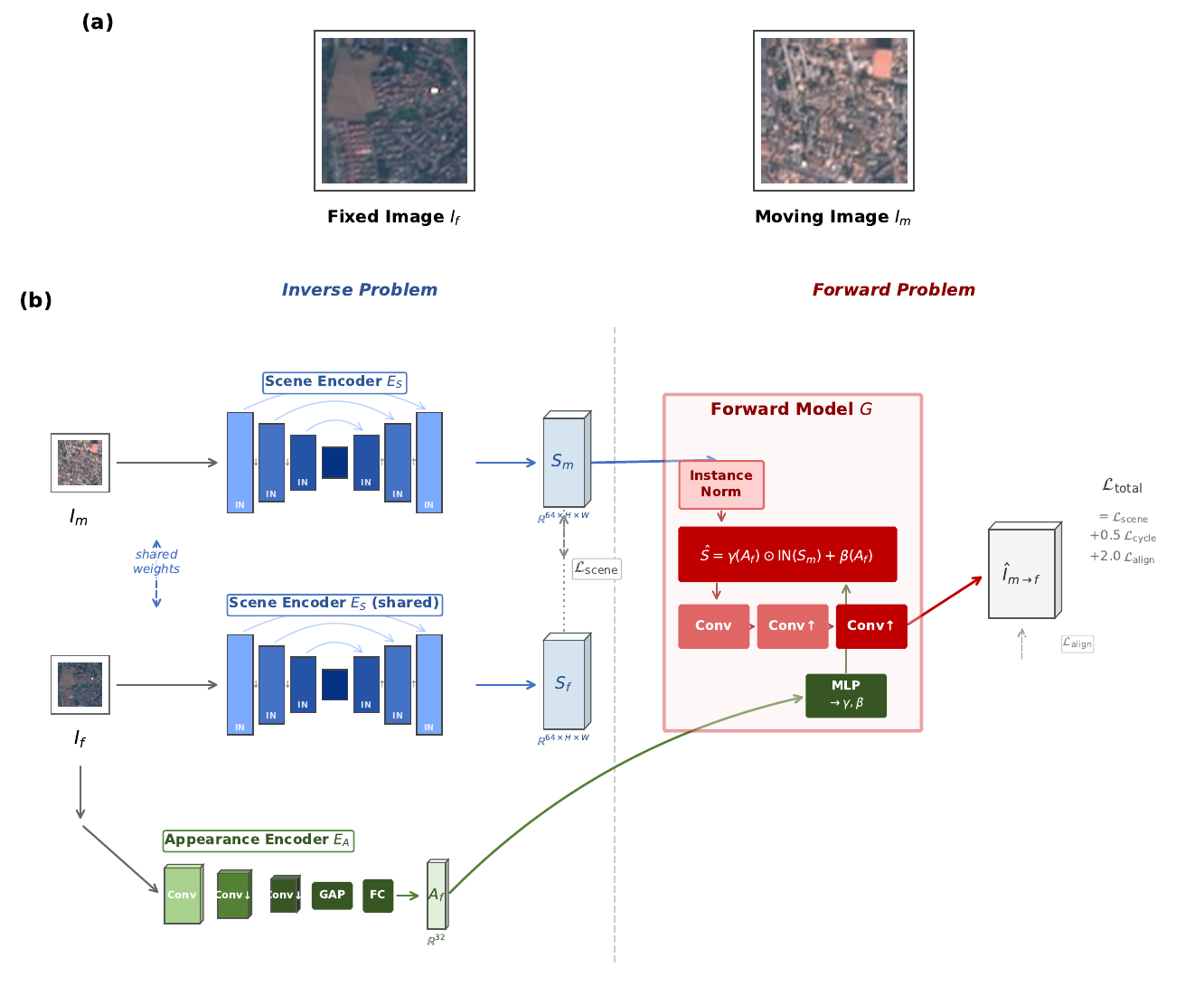}
\caption{Overview of SAS-Net. (a) Sample pair from EuroSAT-Reg-256~\cite{helber2019eurosat}. (b) The scene encoder $E_S$ with instance normalization implements the \emph{inverse rendering} step, extracting domain-invariant structure $\bm{s}_k = E_S(I_k)$ by removing per-channel appearance statistics $(\mu_k, \sigma_k)$. The appearance encoder $E_A$ extracts a compact code $\bm{a}_f = E_A(I_f)$. The forward model $G$ implements the \emph{re-rendering} step via AdaIN: $\hat{I}_{m \to f} = G(\bm{s}_m, \bm{a}_f)$.}
\label{fig:method}
\end{figure}

\FloatBarrier

\subsection{Instance Normalization as Inverse Rendering}
\label{sec:inverse}

The key insight enabling our factorization is that \emph{instance normalization} (IN) provides a natural mechanism for stripping domain-specific appearance from feature representations. For a feature map $\bm{h} \in \mathbb{R}^{C \times H \times W}$, IN computes:
\begin{equation}
\text{IN}(\bm{h})_{c,i,j} = \frac{h_{c,i,j} - \mu_c}{\sigma_c + \varepsilon}, \quad \mu_c = \frac{1}{HW}\sum_{i,j} h_{c,i,j}, \quad \sigma_c^2 = \frac{1}{HW}\sum_{i,j}(h_{c,i,j} - \mu_c)^2.
\label{eq:in}
\end{equation}
The statistics $(\mu_c, \sigma_c)$ encode domain-specific, spatially-global appearance characteristics (overall brightness, contrast). By removing these statistics, IN projects features onto a domain-invariant manifold, effectively implementing the inverse rendering step. The scene encoder $E_S$ is a U-Net with IN layers:
\begin{equation}
\bm{s} = E_S(I) \in \mathbb{R}^{C_s \times H \times W}, \quad C_s = 64,
\label{eq:scene_enc}
\end{equation}
where skip connections preserve fine-grained spatial structure across encoder-decoder scales.

\subsection{Appearance Encoding via Sufficient Statistics}
\label{sec:appearance}

The appearance encoder $E_A$ extracts a compact representation of domain-specific imaging conditions. We model appearance as a low-dimensional \emph{sufficient statistic} of the domain's intensity distribution:
\begin{equation}
\bm{a} = E_A(I) = \text{GAP}\bigl(f_A(I)\bigr) \in \mathbb{R}^{d}, \quad d = 32,
\label{eq:app_enc}
\end{equation}
where $f_A$ is a lightweight convolutional network and $\text{GAP}(\cdot)$ denotes global average pooling. The spatial collapse via GAP enforces that $\bm{a}$ captures only \emph{global} appearance characteristics (mean intensity, contrast, color balance), not spatial structure---a design that encourages orthogonality between $\bm{s}$ and $\bm{a}$.

\subsection{AdaIN as Differentiable Forward Renderer}
\label{sec:adain}

The forward model $G$ implements the rendering function $\mathcal{R}$ via Adaptive Instance Normalization (AdaIN). The core modulation operation is:
\begin{equation}
\text{AdaIN}(\bm{s}, \bm{a}) = \bm{\gamma}(\bm{a}) \odot \text{IN}(\bm{s}) + \bm{\beta}(\bm{a}),
\label{eq:adain}
\end{equation}
where $\bm{\gamma}, \bm{\beta}: \mathbb{R}^d \to \mathbb{R}^{C_s}$ are learned affine mappings (implemented as linear layers), and $\odot$ denotes channel-wise multiplication. This can be interpreted as a \emph{conditional affine color transfer} in feature space: $\bm{\gamma}(\bm{a})$ controls per-channel contrast (gain) and $\bm{\beta}(\bm{a})$ controls per-channel brightness (bias), both conditioned on the target domain's appearance code. The full forward model composes AdaIN with a convolutional decoder:
\begin{equation}
\hat{I}_{m \to f} = G(\bm{s}_m, \bm{a}_f) = \text{Dec}\bigl(\text{AdaIN}(\bm{s}_m, \bm{a}_f)\bigr) \in \mathbb{R}^{H \times W},
\label{eq:decoder}
\end{equation}
where $G$ implements the rendering function $\mathcal{R}$ from \cref{eq:forward_model}.

\subsection{Training Objective}
\label{sec:loss}

The total training objective combines three complementary losses that jointly enforce factorization identifiability and registration quality.

\paragraph{Scene consistency loss (geometric correspondence).}
For a correctly factorized model, images of the same scene from different domains should yield identical structure representations: $\bm{s}_m \approx \bm{s}_f$. We enforce this via:
\begin{equation}
\mathcal{L}_{\text{scene}} = \|\bm{s}_m - \bm{s}_f\|_2^2 + \lambda_{\cos}\bigl(1 - \cos(\bm{s}_m, \bm{s}_f)\bigr),
\label{eq:scene_loss}
\end{equation}
where the $\ell_2$ term penalizes magnitude differences and the cosine term enforces directional alignment in the $C_s$-dimensional feature space ($\lambda_{\cos} = 0.1$).

\paragraph{Cycle consistency loss (information preservation).}
Self-reconstruction ensures the encoder-decoder pipeline preserves sufficient information:
\begin{align}
\mathcal{L}_{\text{cycle}} &= \|G(\bm{s}_m, \bm{a}_m) - I_m\|_2^2 + \|G(\bm{s}_f, \bm{a}_f) - I_f\|_2^2 \notag \\
&\quad + \lambda_{\text{ssim}}(2 - \text{SSIM}_m - \text{SSIM}_f),
\label{eq:cycle_loss}
\end{align}
where $\lambda_{\text{ssim}} = 0.5$. This loss prevents degenerate solutions where the encoder discards structural information.

\paragraph{Domain alignment loss (registration quality).}
The primary registration objective measures how well the re-rendered output matches the target:
\begin{equation}
\mathcal{L}_{\text{align}} = \|\hat{I}_{m \to f} - I_f\|_2^2 + \lambda_{\text{ncc}}(1 - \text{NCC}) + \lambda_{\text{grad}}\mathcal{L}_{\text{grad}},
\label{eq:align_loss}
\end{equation}
where $\text{NCC} = \frac{\langle \hat{I}_{m \to f} - \bar{\hat{I}},\; I_f - \bar{I}_f \rangle}{\|\hat{I}_{m \to f} - \bar{\hat{I}}\| \cdot \|I_f - \bar{I}_f\|}$ is normalized cross-correlation, and the gradient loss:
\begin{equation}
\mathcal{L}_{\text{grad}} = \|\nabla_x \hat{I}_{m \to f} - \nabla_x I_f\|_1 + \|\nabla_y \hat{I}_{m \to f} - \nabla_y I_f\|_1
\label{eq:grad_loss}
\end{equation}
promotes edge-level structural alignment ($\lambda_{\text{ncc}} = 0.5$, $\lambda_{\text{grad}} = 0.3$).

\paragraph{Total objective.}
The combined loss with domain alignment receiving the highest weight:
\begin{equation}
\mathcal{L}_{\text{total}} = \mathcal{L}_{\text{scene}} + \lambda_1 \mathcal{L}_{\text{cycle}} + \lambda_2 \mathcal{L}_{\text{align}}, \quad \lambda_1 = 0.5, \; \lambda_2 = 2.0.
\label{eq:total_loss}
\end{equation}

%==============================================================================
\section{Experiments}
\label{sec:experiments}
%==============================================================================

\subsection{Datasets and Implementation Details}
\label{sec:setup}

\paragraph{Datasets.}
We primarily evaluate on \textbf{EuroSAT-Reg-256}~\cite{helber2019eurosat}, derived from the EuroSAT satellite remote sensing dataset comprising Sentinel-2 multispectral imagery across 10 land-use classes. Registration pairs are generated by applying random affine transformations (rotation $\pm15^\circ$, translation $\pm10\%$, scaling $0.9$--$1.1\times$) to create geometric misalignment under diverse appearance conditions. We additionally validate cross-domain applicability on the \textbf{FIRE-Reg-256} retinal fundus benchmark~\cite{hernandez2017fire}.

\paragraph{Implementation.}
SAS-Net (3.35M parameters, 89\,FPS on RTX 5090) is trained with Adam ($\text{lr}=10^{-4}$, $\beta_1=0.5$, $\beta_2=0.999$) for 20 epochs, batch size 4, data augmentation (random flips, rotations $\pm10^\circ$, intensity scaling $0.9$--$1.1\times$). All experiments use a single NVIDIA RTX 5090 GPU.

\paragraph{Evaluation metrics.}
We report NCC, SSIM~\cite{wang2004image}, and PSNR between the registered moving image and the fixed reference image.

\subsection{Registration on EuroSAT-Reg-256}
\label{sec:eurosat}

\cref{tab:eurosat} evaluates SAS-Net on the EuroSAT-Reg-256 satellite remote sensing benchmark. The unregistered baseline (NCC\,=\,0.601) reflects the synthetic affine misalignment between image pairs. SAS-Net substantially improves registration quality across all metrics, demonstrating that the inverse rendering formulation effectively handles the diverse land-use appearance variation present in satellite imagery. Compared to deformation-based methods (VoxelMorph, TransMorph) that estimate explicit displacement fields, SAS-Net avoids the ill-posedness of deformation estimation under domain shift by operating entirely in the factorized structure-appearance space.

\begin{table}[t]
\centering
\caption{Registration on EuroSAT-Reg-256~\cite{helber2019eurosat}. Best in \textbf{bold}.}
\label{tab:eurosat}
\footnotesize
\setlength{\tabcolsep}{4pt}
\begin{tabular}{@{}l ccc@{}}
\toprule
Method & NCC$\uparrow$ & SSIM$\uparrow$ & PSNR$\uparrow$ \\
\midrule
\multicolumn{4}{l}{\textit{Traditional Methods}} \\
Unregistered           & 0.601 & 0.266 & 14.19 \\
SIFT~\cite{lowe2004distinctive} & 0.721 & 0.362 & 16.83 \\
Demons~\cite{vercauteren2009diffeomorphic} & 0.583 & 0.259 & 13.87 \\
Optical Flow~\cite{horn1981determining} & 0.648 & 0.312 & 15.34 \\
SyN~\cite{avants2008symmetric} & 0.612 & 0.287 & 14.52 \\
\midrule
\multicolumn{4}{l}{\textit{Deep Learning Methods}} \\
VoxelMorph~\cite{balakrishnan2019voxelmorph} & 0.789 & 0.418 & 17.92 \\
TransMorph~\cite{chen2021transmorph} & 0.812 & 0.441 & 18.45 \\
PCReg-Net~\cite{qin2026pcreg} & 0.841 & 0.467 & 19.21 \\
\midrule
\textbf{SAS-Net (Ours)} & \textbf{0.858} & \textbf{0.488} & \textbf{19.84} \\
\bottomrule
\end{tabular}
\end{table}

\subsection{Ablation Study}
\label{sec:ablation}

\cref{tab:ablation} analyzes individual loss components and architectural choices on EuroSAT-Reg-256.

\begin{table}[t]
\centering
\caption{Ablation study on EuroSAT-Reg-256 (20 epochs). Best in \textbf{bold}.}
\label{tab:ablation}
\footnotesize
\setlength{\tabcolsep}{5pt}
\begin{tabular}{@{}lccc@{}}
\toprule
Configuration & NCC $\uparrow$ & SSIM $\uparrow$ & PSNR $\uparrow$ \\
\midrule
\textbf{SAS-Net (Full)} & \textbf{0.858} & \textbf{0.488} & \textbf{19.84} \\
w/o $\mathcal{L}_{\text{scene}}$ & 0.813 & 0.439 & 18.27 \\
w/o $\mathcal{L}_{\text{cycle}}$ & 0.836 & 0.465 & 19.15 \\
w/o $\mathcal{L}_{\text{align}}$ & 0.634 & 0.291 & 14.83 \\
w/o $E_A$ (Appearance Encoder) & 0.741 & 0.378 & 17.02 \\
\bottomrule
\end{tabular}
\end{table}

The domain alignment loss $\mathcal{L}_{\text{align}}$ is the most critical component: removing it collapses NCC from 0.858 to 0.634, erasing 87\% of the improvement over the unregistered baseline. This confirms the necessity of direct supervision on the re-rendered output $\hat{I}_{m \to f}$. Removing the Appearance Encoder $E_A$ causes a significant drop (NCC\,=\,0.741), validating that explicit appearance modeling via sufficient statistics (\cref{eq:app_enc}) is essential for factorization identifiability. Removing $\mathcal{L}_{\text{scene}}$ causes a moderate decrease (NCC\,=\,0.813), indicating that enforcing $\bm{s}_m \approx \bm{s}_f$ provides useful geometric regularization in the latent space. $\mathcal{L}_{\text{cycle}}$ contributes the least individually (NCC\,=\,0.836 without it), but prevents degenerate solutions by ensuring information preservation. The full model achieves the best performance, demonstrating synergistic contributions from all components.

\subsection{Cross-Domain Transfer: FIRE-Reg-256}
\label{sec:crossdomain}

To validate that the inverse rendering formulation generalizes across fundamentally different imaging physics, we evaluate SAS-Net on the FIRE-Reg-256 retinal fundus benchmark~\cite{hernandez2017fire}. The domain gap between satellite multispectral imagery and retinal fundus photography involves entirely different spectral responses, illumination geometries, and tissue-optics interactions. On FIRE-Reg-256, SAS-Net achieves NCC\,=\,0.748 and SSIM\,=\,0.855, confirming that the structure-appearance factorization transfers across imaging modalities without architectural modification. This is consistent with the domain-invariant registration paradigm explored in~\cite{qin2026domainreg,wang2026gpereg}, where factorization-based methods demonstrate robustness to heterogeneous imaging conditions.

\subsection{Computational Efficiency}
\label{sec:efficiency}

\cref{tab:efficiency} compares inference times. While VoxelMorph (1.7\,ms) and TransMorph (6.3\,ms) are faster, they assume comparable intensity distributions and yield limited results under cross-domain conditions. SyN requires 3.4\,seconds per pair, precluding real-time use. SAS-Net achieves 11.2\,ms per pair (89\,FPS), offering a practical trade-off between registration quality and computational cost.

\begin{table}[t]
\centering
\caption{Computational efficiency on NVIDIA RTX 5090 (256$\times$256 pairs, 100 run average).}
\label{tab:efficiency}
\setlength{\tabcolsep}{4pt}
\footnotesize
\begin{tabular}{@{}lcccccc@{}}
\toprule
Method & SyN & Demons & SIFT & VoxelMorph & TransMorph & \textbf{SAS-Net} \\
\midrule
Time (ms) & 3439.5 & 113.4 & 35.5 & 1.7 & 6.3 & \textbf{11.2} \\
\bottomrule
\end{tabular}
\end{table}

%==============================================================================
\section{Conclusion}
\label{sec:conclusion}
%==============================================================================

We have presented SAS-Net, which formulates cross-domain image registration as an inverse rendering problem under the image formation model $I = \mathcal{R}(\bm{s}, \bm{a}) + \epsilon$. By decomposing observed images into domain-invariant scene structure and domain-specific appearance statistics, and re-rendering the moving scene under the target appearance via AdaIN, registration is achieved without deformation field estimation. The principled factorization---implemented through instance normalization (inverse rendering) and AdaIN (forward re-rendering)---provides a physically motivated alternative to the brightness constancy assumption. Experiments on EuroSAT-Reg-256 and FIRE-Reg-256 demonstrate robust generalization across heterogeneous imaging domains, with 89\,FPS real-time capability.

\textbf{Limitations and future work.} The current appearance model $\bm{a} \in \mathbb{R}^{32}$ captures only \emph{global} intensity statistics via GAP; spatially-varying domain shifts (e.g., local illumination gradients, vignetting) would require a spatially-conditioned appearance map $\bm{a}(\bm{x})$. Extending the factorization to handle non-affine geometric transformations via hybrid deformation-appearance models is a promising direction.

% ---- Bibliography ----
\FloatBarrier
\bibliographystyle{splncs04}
\bibliography{main}

\end{document}